\documentclass[aps,prl,superscriptaddress,twocolumn]{revtex4-1} % for PRL format
\usepackage{amsmath}
\usepackage{epstopdf}
\usepackage{graphicx}
\begin{document}

\preprint{}

\title{Violation of Continuous Variable EPR Steering with Discrete Measurements}


\author{James Schneeloch}
\affiliation{Department of Physics and Astronomy, University of Rochester, Rochester, NY 14627}

\author{P. Ben Dixon}
\affiliation{Research Laboratory of Electronics, Massachusetts
Institute of Technology, Cambridge, MA 02139}
\author{Gregory A. Howland}
\affiliation{Department of Physics and Astronomy, University of Rochester, Rochester, NY 14627}
\author{Curtis J. Broadbent}
\affiliation{Department of Physics and Astronomy, University of Rochester, Rochester, NY 14627}
\affiliation{Rochester Theory Center, University of Rochester, Rochester, NY 14627}
\author{John C.  Howell}
\affiliation{Department of Physics and Astronomy, University of Rochester, Rochester, NY 14627}

%\email[]{jschneel@pas.rochester.edu, curtis@pas.rochester.edu}
%\thanks{}
%\affiliation{University of Rochester, Dept. of Physics and Astronomy}


\date{\today}

\begin{abstract}
In this Letter, we derive an entropic Einstein-Podolsky-Rosen $($EPR$)$ steering inequality for continuous variable $($CV$)$ systems using only experimentally measured discrete probability distributions and details of the measurement apparatus. We use this inequality to witness EPR steering between the positions and momenta of photon pairs generated in spontaneous parametric downconversion $($SPDC$)$. We examine the asymmetry between parties in this inequality, and show that this asymmetry can be used to reduce the technical requirements of experimental setups intended to demonstrate the EPR paradox. Furthermore, we develop a more stringent steering inequality that is symmetric between parties, and use it to show that the downconverted photon pairs also exhibit symmetric EPR steering.
\end{abstract}

\pacs{03.67.Mn, 03.67.-a, 03.65-w, 42.50.Xa}
%\keywords{}

\maketitle


Witnessing EPR steering \cite{Wiseman2007} is an intuitive way to demonstrate the dichotomy between completeness and local realism in the EPR paradox  \cite{EPR1935}, a task whose difficulty grows with the dimensionality of the system. Violating EPR steering for continuous variables $($CV$)$ is understandably challenging because CV systems are infinite-dimensional. As interest in EPR steering increases, including interest regarding its fundamental aspects \cite{PhysRevA.86.010101,Walborn2011,Dixon2012,Reid1989,Howell2004,Edgar2012,Handchen2012}, as well as its use in quantum information applications such as secure quantum key distribution \cite{BranciardQKD2012}, witnesses of CV EPR steering become increasingly important.

EPR steering, first formulated by Wiseman \emph{et~al.}~\cite{Wiseman2007} embodies a level of quantum correlation weaker than Bell-nonlocality \cite{CHSHbell1969}, but stronger than mere nonseparability \cite{Peres1996}. Consider a maximally entangled pair of particles A and B. By choosing an observable to measure on A, one can, ``steer'', B to be well-defined in that same observable $($whatever it is$)$ without directly interacting with it. It is this nonlocal influence that is captured in EPR steering.

To show that A can steer B by these nonlocal means, the correlations between particles A and B must be strong enough to rule out any model of local hidden states (LHS) from describing B. This occurs when the conditional or inferred measurement outcomes of B no longer obey the same uncertainty relations that a single particle does \cite{Cavalcanti2009}. In such a case, we demonstrate EPR's dichotomy, that either quantum mechanics offers an incomplete description of B, or we must allow that these correlations can be nonlocal in nature.

Reid \cite{Reid1989} was the first to create an EPR steering inequality using the Heisenberg uncertainty principle. Later, Walborn \emph{et~al.}~\cite{Walborn2011} developed a steering inequality using an entropic \cite{Shannon1949} formulation of the uncertainty principle for CV position and momentum \cite{BiałynickiBirula1975}.  For all states whose spatial correlations are insufficient to demonstrate the EPR paradox, the inequality,
\begin{equation}\label{WSE}
h(\vec{x}_{B}|\vec{x}_{A}) + h(\vec{k}_{B}|\vec{k}_{A})\geq n \log(\pi e),
\end{equation}
must be satisfied where: $n$ is the number of spatial dimensions; $h(\vec{x}_{B}|\vec{x}_{A})$ is the continuous Shannon entropy \cite{Cover2006} of the distribution of measurement outcomes of the position of particle $B$, $\vec{x}_{B}$, conditioned on the measurement outcomes of the position of particle $A$, $\vec{x}_{A}$; $h(\vec{k}_{B}|\vec{k}_{A})$ is similarly defined for measurements of the wavenumber or momentum in natural units; and the base of the logarithm is determined by the units in which we choose to measure the entropy. Walborn \emph{et~al.}~proved this inequality for one dimension, but it is easily generalized to $n$ dimensions assuming that different spatial degrees of freedom are statistically independent. 

Bialynicki-Birula and Mycielski derived single-particle CV entropic uncertainty relations \cite{BiałynickiBirula1975}.  They showed that these uncertainty relations were stronger than any variance-based uncertainty relation for all quantum states and were only equal for minimum-uncertainty Gaussian states.   For this reason, Walborn's entropic steering inequality is stronger than variance-based steering inequalities \cite{Reid1989, Reid2009} and equal only for minimum-uncertainty Gaussian states. As powerful as Walborn's entropic steering inequality \eqref{WSE} is, it cannot be used in the laboratory because continuous probability densities cannot be determined with a finite number of measurements.

In this Letter, we derive an entropic EPR steering inequality suitable for experimental investigations of CV position-momentum entanglement with discrete measurements:
\begin{equation}\label{discWSE}
H(\vec{X_{B}}|\vec{X}_{A}) + H(\vec{K}_{B}|\vec{K}_{A}) \geq \sum_{i=1}^{n} \log \bigg(\frac{\pi e}{\Delta x_{Bi} \Delta k_{Bi}} \bigg).
\end{equation}
Here we have that $H(\vec{X}_{B}|\vec{X}_{A})$ is the discrete Shannon entropy \cite{Cover2006} of measurements of the position of particle B conditioned on measurements of the position of particle A, where both position domains have been discretized into equally spaced windows reflecting the precision of the experimental setup. Though the measurements are discrete, this inequality witnesses EPR steering in continuous position and momentum without needing to determine probability density functions. In addition, we note that the position and momentum referred to in this paper are not field quadratures, but actual positions and momenta of downconverted photons as seen in Ref.~\cite{Dixon2012}.

Our inequality \eqref{discWSE} provides both a simpler and more powerful method to successfully witness EPR steering. We do not need to reconstruct CV probability density functions, and we only require experimentally resolvable discrete probabilities along with those details of the experimental setup needed to determine the measurement resolutions $\Delta x_{Bi}$ and $\Delta k_{Bi}$. As we will show, violation of our steering inequality \eqref{discWSE} also violates the steering inequality derived by Walborn \emph{et~al.} \eqref{WSE}, which is the most powerful CV EPR steering inequality to date. Since ours can be used in the lab, this represents a significant advance in our ability to experimentally witness EPR steering in states which couldn't be witnessed otherwise.

To derive our inequality, we will use a fundamental connection between continuous and discrete entropies \eqref{fundamentalconnection} studied in Ref.~\cite{Rudnicki2012}, to show that any two continuous random variables $x$ and $y$ that can be discretized into equally spaced windows of size $\Delta x$ and $\Delta y$ satisfy the following inequality;
\begin{equation} \label{myineq1}
h(y|x) \leq H(Y|X) + \log(\Delta y).
\end{equation} 

Consider an experiment to measure random variable $x$ which can take the value of any real number with probability density $\rho(x)$. The experiment is only capable of measuring $x$ to discrete windows $X_{\ell}$ of size $\Delta x$. The probability of measuring $x$ to be in window $X_{\ell}$ is
\begin{equation}\label{discprob}
P(X_{\ell}) \equiv \int\limits_{\Delta x_{\ell}} dx\;  \rho(x),
\end{equation}
where the region of integration $\Delta x_{\ell}$ is the range of values of $x$ between $x_{\ell} - \frac{1}{2}\Delta x$ and  $x_{\ell} + \frac{1}{2}\Delta x$, and $x_{\ell}$ is the value of $x$ at the center of the window $X_{\ell}$; i.e. $x$ is subdivided into equal-size segments $X_{\ell}$ of size $\Delta x$, and $\Delta x_{\ell}$ is the range of values of $x$ in $X_{\ell}$. The Shannon entropy of this discrete probability distribution is given by
\begin{equation}
H(X) = - \sum_{\ell} P(X_{\ell}) \log(P(X_{\ell})),
\end{equation}
and the Shannon entropy \cite{Cover2006} of the continuous probability density function $\rho(x)$ is expressed as
\begin{equation}
h(x) = - \int dx\;\rho(x)\log(\rho(x)).
\end{equation}
We now define the distribution $\rho_{\ell}(x)$ as the probability distribution of $x$ conditioned on being measured within window $X_{\ell}$. The continuous entropy $h_{\ell}(x)$ is defined as the entropy of $\rho_{\ell}(x)$ where
\begin{equation}
\rho_{\ell}(x) = \frac{\rho(x)}{P(X_{\ell})}
\end{equation}
for all values of $x$ in the window $X_{\ell}$, and is zero otherwise.  By breaking up the continuous entropy $h(x)$ into a sum over all windows, and expressing $h(x)$ in terms of $h_{\ell}(x)$ and $P(X_{\ell})$, we obtain the fundamental connection between discrete and continuous entropies;
\begin{equation}\label{fundamentalconnection}
h(x) = \sum_{\ell} P(X_{\ell}) h_{\ell}(x) + H(X).
\end{equation}
This connection exists for joint entropies as well as marginal entropies, where we now define $h_{\ell m}(x,y)$ as the entropy of the joint distribution $\rho_{\ell m}(x,y)$ conditioned on $x$ being measured within window $X_{\ell}$ and $y$ being measured within window $Y_{m}$.

The conditional entropies $h(y|x)$ and $H(Y|X)$ are defined as differences between joint and marginal entropies \cite{Cover2006},
\begin{subequations}
\begin{align}\label{def}
h(y|x)&\equiv h(x,y) - h(x),\\
H(Y|X)&\equiv H(X,Y) - H(X).
\end{align}
\end{subequations}
By using \eqref{fundamentalconnection} for both single and joint entropies and \eqref{def} and the knowledge that conditioning on additional events reduces the average entropy, it can be shown that
\begin{equation}
h(y|x) \leq  \sum_{\ell,m} P(X_{\ell},Y_{m}) h_{\ell m}(y|x) + H(Y|X).
\end{equation}
The uniform distribution maximizes the entropy \cite{Cover2006}, so that when all windows $\Delta y_{m}$ are of equal size, we have $h_{\ell m}(y|x)\leq \log(\Delta y)$, which completes our proof of \eqref{myineq1}. Where $x_{Ai}$ and $x_{Bi}$ are another ordinary pair of random variables, we can substitute the expression ~\eqref{myineq1} into the steering inequality created by Walborn \emph{et~al.}~\eqref{WSE} to derive our entropic EPR steering inequality suitable for experimental investigations of CV entanglement. For a particular spatial degree of freedom $i\in\{1,...,n\}$, $($i.e. a particular dimension in space$)$ we've shown that
\begin{equation}\label{discWSE2}
H(X_{Bi}|X_{Ai}) + H(K_{Bi}|K_{Ai}) \geq \log \bigg(\frac{\pi e}{\Delta x_{Bi} \Delta k_{Bi}} \bigg).
\end{equation}
When different spatial degrees of freedom are statistically independent of one another, the entropies add, giving us the $n$-dimensional discrete steering inequality we sought to prove \eqref{discWSE}.

When applying our steering inequality \eqref{discWSE}, there are a number of critical details to consider. First, our discrete steering inequalities for each degree of freedom \eqref{discWSE2} have a cutoff of experimental resolution below which our ability to probe quantum phenomena dependent on the uncertainty principle ceases to exist. For any $\Delta x_{Bi} \Delta k_{Bi}$ larger than $\pi e$, we have insufficient resolution to witness entanglement in the \emph{i}$^{th}$ degree of freedom with our steering inequality \eqref{discWSE2}. In this regime, the steering inequality for the \emph{i}$^{th}$ degree of freedom \eqref{discWSE2} becomes impossible to violate since the bound on the right hand side becomes negative, while the sum of discrete entropies on the left hand side is always nonnegative. However,  as we increase the resolution, decreasing $\Delta x_{Bi} \Delta k_{Bi}$, we are better able to violate our EPR steering inequality.

\begin{figure}[h]
\includegraphics[width=\columnwidth,height = 0.3\textheight]{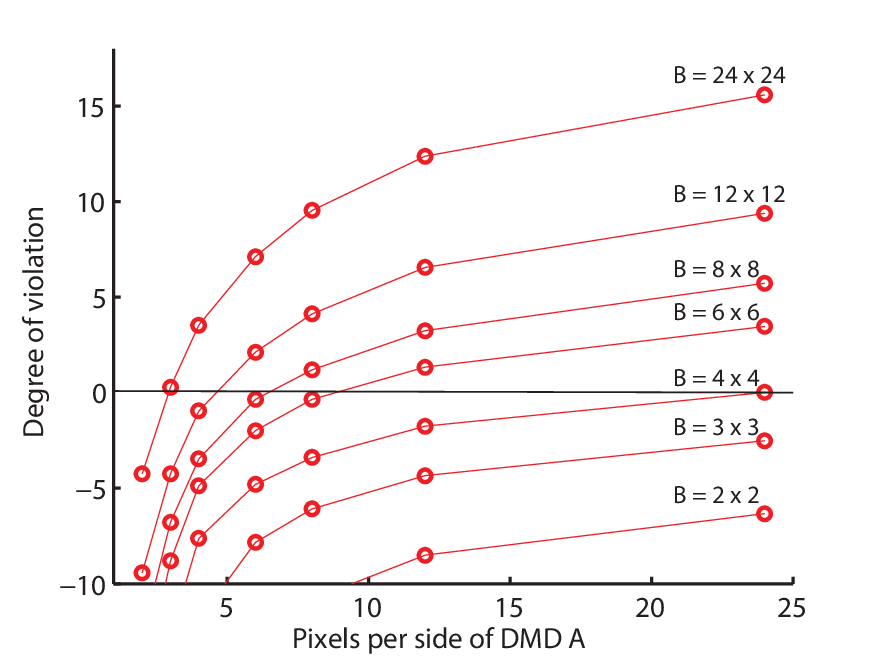}
\caption{Violation of our EPR steering inequality \eqref{discWSE} for measurements of party B conditioned on measurements of party A as measured in standard deviations of the difference between the bound and the sum of conditional entropies. This plot was formed by independently downsampling the joint distributions for a $24\times24$ data set. A positive value represents a violation. We see that below $4\times4$ resolution for party B as suggested by theory, the inequality is not violated for any resolution of DMD A, but violation is possible for large resolution of party B and party A having resolution below $4\times4$.}
\label{fig1}
\end{figure}

Secondly, the inequality \eqref{discWSE} appears to depend on the resolutions of only one detector. Though the conditional entropies have an inherent dependence on the resolution of both parties, the inequality is  asymmetric, as we show in Fig.~\ref{fig1} with experimental data from Ref.~\cite{Dixon2012}.

Using Bayes' rule to swap parties \cite{Cover2006}, we see that symmetry between parties exists only when the marginal entropies for each party are equal to one another. Due to the asymmetry between parties in \eqref{discWSE}, it is not true in general that the ability to violate the discrete inequality on a given subsystem is simply limited by the party with lowest resolution. However, since the double gaussian wavefunction used to model SPDC is symmetric between parties \cite{Dixon2012}, experimental investigations of EPR steering in SPDC do not exhibit a high degree of asymmetry, though the asymmetry is still significant enough to reduce the necessary technical specifications of an experiment to witness EPR steering. Particularly at higher resolutions, one detector can have less than half the resolution of the other and still be used to witness steering.

In addition, we consider the case where the product of position and momentum windows is fixed, but each is allowed to vary. For large position window sizes: the position conditional entropy decreases toward zero, and the momentum conditional entropy increases without limit. This makes the sum of discrete entropies in \eqref{discWSE} grow without limit while the uncertainty bound remains constant, making EPR steering progressively more difficult to observe. Downsampling the momentum distribution would not make the inequality easier to violate because doing so would also decrease the bound on the right hand side even though it would result in a smaller discrete entropy on the left hand side of the inequality.

Expressing our steering inequality in terms of conditional entropies is useful, but there is also a stronger level of entanglement  $($which we call symmetric EPR steering$)$ sufficient to allow EPR steering between both parties. This level of entanglement can be witnessed by expressing our inequality in terms of the mutual information.
The mutual information \cite{Cover2006} is defined as
\begin{equation}
I(\vec{X}_{A},\vec{X}_{B}) \equiv H(\vec{X}_{A}) + H(\vec{X}_{B}) - H(\vec{X}_{A},\vec{X}_{B}).
\end{equation}
Our steering inequality \eqref{discWSE} expressed in terms of mutual information becomes
\begin{align}
I(\vec{X}_{A},\vec{X}_{B}) &+ I(\vec{K}_{A},\vec{K}_{B}) \leq \sum_{i=1}^{n} \log \bigg(\frac{\Delta x_{Bi} \Delta k_{Bi}}{\pi e} \bigg) \nonumber \\
 + &( H(\vec{X}_{B}) + H(\vec{K}_{B}) ).
\end{align}
Since the entropies under discussion are discrete, they are bounded above by the logarithm of the number of windows in the viewing area which can be expressed in terms of ratios $\frac{L_{xi}L_{ki}}{\Delta x_{Bi} \Delta k_{Bi}}$, where $L_{xi}$ and $L_{ki}$ are defined as the total extent of the viewing area in the \emph{i}$^{th}$ direction for position and momentum measurements of party B, respectively. Canceling out like terms and taking the maximum value for the bound between parties A and B, we arrive at a more restrictive steering inequality which is symmetric between parties;
\begin{equation}\label{mutinfoineq}
I(\vec{X}_{A},\vec{X}_{B}) + I(\vec{K}_{A},\vec{K}_{B}) \leq \max_{A,B} \log\bigg(\frac{\prod_{i=1}^{n} L_{xi} L_{ki}}{(\pi e)^{n}}\bigg).
\end{equation}
Violation of this symmetric steering inequality simultaneously witnesses EPR steering for both parties; no conditional distribution of measurement outcomes can be ascribed to a single quantum state.

%\section{Experimental Implementation}

To use these inequalities in practice, we used data from the experiment in Ref.~\cite{Dixon2012} where we performed measurements of the near field and far field probability distributions $($positions and momenta$)$ of pairs of entangled photons generated in SPDC by assembling histograms of coincidence counts.  In Ref.~\cite{Dixon2012}, we were able to use this joint probability distribution to calculate the conditional entropy and the mutual information, but not to determine whether the system had entanglement. With our new EPR steering inequalities, we can verify that the system in Ref.~\cite{Dixon2012} at certain resolutions does exhibit both EPR and symmetric EPR steering, and therefore is not only entangled, but sufficiently entangled to demonstrate the EPR paradox.

In order to better explain what was measured, we provide a brief description of the experimental setup in Ref.~\cite{Dixon2012} which generated the joint probability distribution that we analyze here with our discrete steering inequalities \eqref{discWSE} and\eqref{mutinfoineq}. We separated the downconverted photons with a 50:50 beamsplitter into signal and idler arms, and measured coincident detections using time correlated single photon counting. With these coincidence counts, we measured the joint transverse spatial probability distributions of the photon pairs by imaging the face of the nonlinear crystal (and then its Fourier transform for the momentum correlations) onto DMD (Digital Micromirror Device) arrays in each arm which allowed us to look at both spatial degrees of freedom in the transverse plane. Uncertainties in the probability distribution were estimated by assuming Poissonian statistics. We measured the joint probability distributions at a variety of different resolutions between $8\times8$ pixels and $24\times24$ pixels with the same total viewing area. Primary sources of error were due to the temperature instability of the nonlinear crystal over the time scales needed to take data, the imperfect alignment of the DMD arrays, and the imperfect in-coupling of light from the DMD arrays into our photodetectors.

With the joint probability distributions obtained from measurements for both the positions and momenta of the photon pairs, we calculated the conditional entropies that go into our steering inequality, and used the details of the experimental setup to determine all the window sizes $\Delta x_{Bi}$ and $\Delta k_{Bi}$. Given the details of the experimental setup, and that the two transverse degrees of freedom are considered independent of one another, the discrete steering inequality \eqref{discWSE} takes the form
\begin{align}\label{dixonineq}
 H (\vec{X}_{B}|\vec{X}_{A}) +  & H(\vec{K}_{B}|\vec{K}_{A}) \geq 2 \log\bigg(\frac{\pi e}{\Delta x_{B} \Delta k_{B}}\bigg),
\end{align}
where $\Delta k_{B}$ is the resolution of the detector of party B to distinguish differences in the horizontal component of the momentum. 

\begin{figure}[h]
\includegraphics[width=0.8\columnwidth,height = 0.25\textheight]{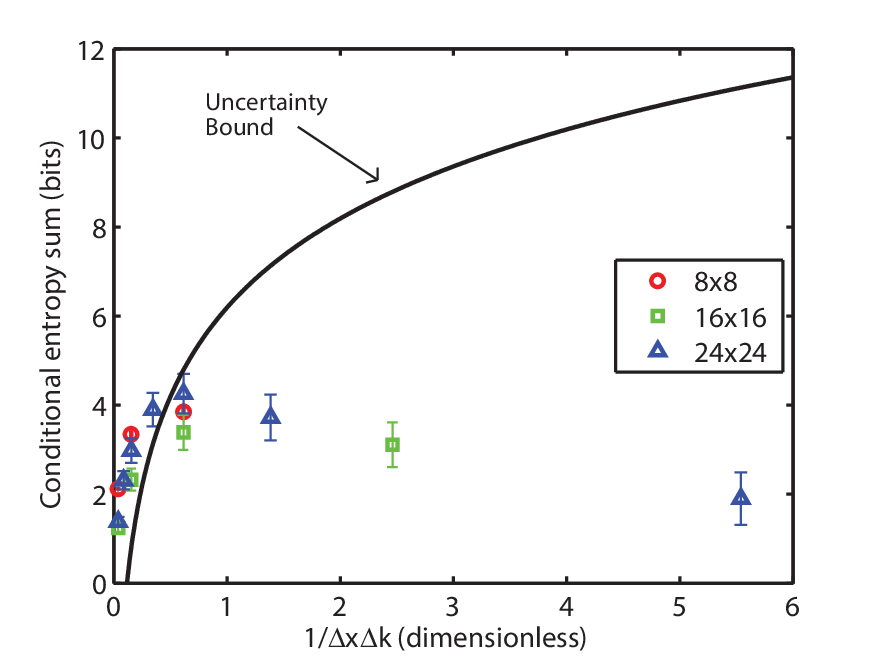}
\caption{Experimentally determined sums of conditional entropies in our EPR steering inequality \eqref{dixonineq} as a function of the product of resolutions $\frac{1}{\Delta x}$ and $\frac{1}{\Delta k}$. The different symbols designate different resolutions of experimental data which were then downsampled. For example, the triangles represent data from an experiment recording at $24\times24$ resolution. Since the product of the window sizes is dimensionless, this can be expressed in any units we wish. The region below the black curve is the region where our EPR steering inequality \eqref{dixonineq} is violated; the sum is less than the bound.}
\label{fig2}
\end{figure}

The experimental data from Ref.~\cite{Dixon2012} relevant to this inequality is displayed in Fig.~\ref{fig2}. In our setup, we were able to violate this inequality by between 3.6 and 16.4 standard deviations for $8\times8$ resolution to $24\times24$ resolution respectively. The values of $L_{x}$ and $L_{k}$ used to determine $\Delta k_{B}$ and $\Delta x_{B}$ were $1.04\times 10^{-3}m$ and $1.00\times 10^{5}m^{-1}$, respectively. The symmetric steering inequality \eqref{mutinfoineq} was not violated for $8\times8$ resolution, but was violated by between 3.4 and 7.0 standard deviations for $16\times16$ resolution and by between  6.6 and 10.7 standard deviations for $24\times24$ resolution. For this type of experiment, noise sources that degrade the measurements and thus prevent quantum correlations from being observed include fluorescence of optics in the system, shot noise of the optical beam, detector noise, and uncertainty in detector pixel size \cite{Brida2009}. In our particular system with high quality optics, low noise detectors, and low light levels, shot noise was the dominant noise source.

%\section{Conclusion}
In this Letter, we derive two new EPR steering inequalities $($\eqref{discWSE} and \eqref{mutinfoineq}$)$ especially well-suited for experimental investigations of CV EPR steering. Our first inequality is more inclusive and with fewer complications than Reid's inequality \cite{Reid1989} at sufficiently high resolution, while both can be used in experiments in contrast to the inequality derived by Walborn \emph{et~al.}~\eqref{WSE} which is more well suited for theoretical analysis. We have successfully witnessed both EPR steering and symmetric EPR steering with experimental data from Ref.~\cite{Dixon2012}, and showed that there is a demonstrable asymmetry between parties which allows steering to be witnessed even when one detector has comparatively low resolution provided the other is sufficiently high. These inequalities are powerful and effective characterization tools that we expect to be widely used in applications ranging from future quantum communication networks to fundamental physical experiments involving high-dimensional quantum states.

\emph{Note added:} Upon completion of this work, we became aware of similar results obtained without using steering inequalities \cite{Tasca2012} . 

We gratefully acknowledge careful editing from Gerardo Viza,  V. Jane Schneeloch, Kyle B. Odell, and Will Bock, as well as support from DARPA DSO InPho Grant No.~W911NF-10-1-0404. CJB acknowledges support from ARO W911NF-09-1-0385 and NSF PHY-1203931.

\subsection{}
\subsubsection{}

\bibliography{EPRbib8}

\end{document}